\documentclass[twocolumn,twoside]{article}
\makeatletter\if@twocolumn\PassOptionsToPackage{switch}{lineno}\else\fi\makeatother

\usepackage{amsmath,amssymb,amstext,tabulary,graphicx,times,caption,fancyhdr}
\usepackage[utf8]{inputenc}
\usepackage[T1]{fontenc}
\usepackage{abstract,booktabs}
\usepackage{url,multirow,morefloats,floatflt,cancel,tfrupee}
\makeatletter

\AtBeginDocument{\@ifpackageloaded{textcomp}{}{\usepackage{textcomp}}}
\makeatother
\usepackage{colortbl}
\usepackage{xcolor}
\usepackage{pifont}
\usepackage[nointegrals]{wasysym}
\makeatletter

\def\mcWidth#1{\csname TY@F#1\endcsname+\tabcolsep}

\def\cAlignHack{\rightskip\@flushglue\leftskip\@flushglue\parindent\z@\parfillskip\z@skip}
\def\rAlignHack{\rightskip\z@skip\leftskip\@flushglue \parindent\z@\parfillskip\z@skip}

\@ifundefined{etal}{}{}

\usepackage{ifxetex}
\ifxetex\else\if@twocolumn\@ifpackageloaded{stfloats}{}{\usepackage{dblfloatfix}}\fi\fi

\AtBeginDocument{
\expandafter\ifx\csname eqalign\endcsname\relax
\def\eqalign#1{\null\vcenter{\def\\{\cr}\openup\jot\m@th
  \ialign{\strut$\displaystyle{##}$\hfil&$\displaystyle{{}##}$\hfil
      \crcr#1\crcr}}\,}
\fi
}

\AtBeginDocument{%
  \@ifpackageloaded{endfloat}%
   {\renewcommand\efloat@iwrite[1]{\immediate\expandafter\protected@write\csname efloat@post#1\endcsname{}}}{\newif\ifefloat@tables}%
}%

\def\BreakURLText#1{\@tfor\brk@tempa:=#1\do{\brk@tempa\hskip0pt}}
\let\lt=<
\let\gt=>
\def\processVert{\ifmmode|\else\textbar\fi}

\@ifundefined{subparagraph}{
\def\subparagraph{\@startsection{paragraph}{5}{2\parindent}{0ex plus 0.1ex minus 0.1ex}%
{0ex}{\normalfont\small\itshape}}%
}{}

\newcommand\role[1]{\unskip}
\newcommand\aucollab[1]{\unskip}
  
\@ifundefined{tsGraphicsScaleX}{\gdef\tsGraphicsScaleX{1}}{}
\@ifundefined{tsGraphicsScaleY}{\gdef\tsGraphicsScaleY{.9}}{}
\def\checkGraphicsWidth{\ifdim\Gin@nat@width>\linewidth
	\tsGraphicsScaleX\linewidth\else\Gin@nat@width\fi}

\def\checkGraphicsHeight{\ifdim\Gin@nat@height>.9\textheight
	\tsGraphicsScaleY\textheight\else\Gin@nat@height\fi}

\def\fixFloatSize#1{}
\let\ts@includegraphics\includegraphics

\def\inlinegraphic[#1]#2{{\edef\@tempa{#1}\edef\baseline@shift{\ifx\@tempa\@empty0\else#1\fi}\edef\tempZ{\the\numexpr(\numexpr(\baseline@shift*\f@size/100))}\protect\raisebox{\tempZ pt}{\ts@includegraphics{#2}}}}

\AtBeginDocument{\def\includegraphics{\@ifnextchar[{\ts@includegraphics}{\ts@includegraphics[width=\checkGraphicsWidth,height=\checkGraphicsHeight,keepaspectratio]}}}

\DeclareMathAlphabet{\mathpzc}{OT1}{pzc}{m}{it}

\def\URL#1#2{\@ifundefined{href}{#2}{\href{#1}{#2}}}

\def\UrlOrds{\do\*\do\-\do\~\do\'\do\"\do\-}%
\g@addto@macro{\UrlBreaks}{\UrlOrds}

\edef\fntEncoding{\f@encoding}

\makeatother

\newif\ifmultipleabstract\multipleabstractfalse%
\usepackage[paperheight=11.02in,paperwidth=8.27in,margin=1.5cm,headsep=.5cm,bottom=1.9cm,top=2.5cm]{geometry}
\usepackage[hang,flushmargin]{footmisc}

 \date{}
\makeatletter\def\bmjIndent{1pt}
\let\@journalTitle\@empty%
\def\journalTitle#1{\gdef\@journalTitle{#1}}

\def\author#1{\gdef\@author{\hskip-\dimexpr(\tabcolsep)\hskip\bmjIndent\parbox{\dimexpr\textwidth-\bmjIndent}{\raggedright\sffamily#1}}}

\def\title#1{\gdef\@title{\raggedright\bfseries\ifx\@articleType\@empty\else\@articleType\\\fi\sffamily#1}}

\let\@articleType\@empty \def\articletype#1{\gdef\@articleType{{\normalfont\itshape#1}}}

\fancypagestyle{headings}{\fancyhf{}\fancyhead[RO]{}\fancyhead[LE]{\ifx\#journalTitle\@empty\else\sffamily\footnotesize\@journalTitle\fi}\fancyfoot[RO,LE]{\sffamily\thepage}\fancyfoot[LO,RE]{\sffamily\footnotesize\RunningAuthor}}
\fancypagestyle{plain}{\fancyhf{}\fancyfoot[R]{\sffamily\thepage}\fancyfoot[C]{\sffamily\footnotesize\RunningAuthor}}

\def\NormalBaseline{\def\baselinestretch{1.1}}

\usepackage[noindentafter]{titlesec}
\titleformat{\section}[hang]{\NormalBaseline\filright\large\bfseries\sffamily}
{\large\thesection}
{10pt}
{\noindent\MakeUppercase}
[]
\titleformat{\subsection}[hang]{\NormalBaseline\filright\bfseries\sffamily}
{\thesubsection}
{10pt}
{}
[]
\titleformat{\subsubsection}[hang]{\NormalBaseline\filright\sffamily}
{\upshape\thesubsubsection}
{10pt}
{}
[]
\titleformat{\paragraph}[runin]{\NormalBaseline\filright\itshape\sffamily}
{\theparagraph}
{10pt}
{}
[]
\titleformat{\subparagraph}[runin]{\NormalBaseline\filright\sffamily}
{\thesubparagraph}
{10pt}
{}
[]
\renewcommand\footnotemark{}
\titlespacing{\section}{0pt}{1.5\baselineskip}{.2\baselineskip}
\titlespacing{\subsection}{0pt}{1.5\baselineskip}{.2\baselineskip}
\titlespacing{\subsubsection}{0pt}{1.5\baselineskip}{.2\baselineskip}
\titlespacing{\paragraph}{0pt}{.5\baselineskip}{10pt}
\titlespacing{\subparagraph}{0pt}{.5\baselineskip}{10pt}

\makeatother

\usepackage[superscript,biblabel]{cite}
\usepackage[]{natbib}

\def\truncatedAt{1000}
\def\typesetArtId{5882794d-bbd9-4478-b160-506dfa394c10}    
    \makeatletter
    \@ifpackageloaded{hyperref}{}{\usepackage[pdfborder={0 0 0},bookmarks=false]{hyperref}}
    \@ifpackageloaded{hyperref}{}{}%
    \def\putUpgradeInfoBox{\@ifundefined{truncatedAt}{\def\truncatedAt{1000}}{}
    \def\up@width@one{\if@twocolumn .95\columnwidth\else .65\columnwidth\fi}%
    \def\up@width@two{\if@twocolumn .5\columnwidth\else .3\columnwidth\fi}%
    \vskip 2pc\nopagebreak
    \noindent\vbox{\centering%
    {\fontfamily{phv}\selectfont\footnotesize\color{blue}%
    	\ifx\typesetArtId\empty%
      	\href{https://typeset.io/documents}{\underline{Edit this article on \smash{Typeset}}}%
      \else%
	    	\href{https://typeset.io/edit/\typesetArtId}{\underline{Edit this article on \smash{Typeset}}}%
      \fi%
      }\\[6pt]
    \par\href{https://www.typeset.io/pricing/?source=upgrade-from-pdf}{\includegraphics[width=\up@width@one]{upgrade-logo-11Nov19.png}}\\[2pt]%
    \par%
    \fontfamily{phv}\selectfont\footnotesize\color{blue}%
    \href{https://typeset.io/}{\underline{www.\smash{typeset}.io}}%
    ~~\,\textcolor{black}{\vrule height 8pt width .7pt}~\,~%
    \href{https://www.typeset.io/orders/coupon/ZfAB8STU/?source=pricing-page-student-discount}{\underline{\smash{Looking for a Discount?}}}%
    }%
    }
    \makeatother
    
\begin{document}

\nocite{*}

\thispagestyle{plain}

\title{Evaluating ChatGPT text-mining of clinical records for obesity monitoring}
\author{Ivo S Fins\textsuperscript{1},
            Heather Davies\textsuperscript{1},
            Sean Farrell\textsuperscript{2},
            Jose R Torres\textsuperscript{3},
            Gina Pinchbeck\textsuperscript{1},
            Alan D Radford\textsuperscript{1} and
            Peter-John Noble\textsuperscript{1}~\\[-3pt]\normalsize\normalfont ~\\
\textsuperscript{\sffamily 1}{\sffamily Small Animal Veterinary Surveillance Network, Institute of Infection, Veterinary and Ecological Sciences\unskip, University of Liverpool\unskip, Liverpool\unskip, UK}~\\
\textsuperscript{\sffamily 2}{\sffamily Department of Computer Science\unskip, Durham University\unskip, Durham\unskip, UK}~\\
\textsuperscript{\sffamily 3}{\sffamily Institute for Animal Health and Food Safety\unskip, University of Las Palmas de Gran Canaria\unskip, Canary Archipelago\unskip, Las Palmas\unskip, Spain}}
\def\RunningAuthor{Ivo S Fins., \textit{et~al.}}

\maketitle 
\section*{ABSTRACT}
Background: Veterinary clinical narratives remain a largely untapped resource for addressing complex diseases. Here we compare the ability of a large language model (ChatGPT) and a previously developed regular expression (RegexT) to identify overweight body condition scores (BCS) in veterinary narratives.

Methods: BCS values were extracted from 4,415 anonymised clinical narratives using either RegexT or by appending the narrative to a prompt sent to ChatGPT coercing the model to return the BCS information. Data were manually reviewed for comparison.

Results: The precision of RegexT was higher (100\%, 95\% CI 94.81{\textendash}100\%) than the ChatGPT (89.3\%; 95\% CI82.75-93.64\%). However, the recall of ChatGPT (100\%. 95\% CI 96.18-100\%) was considerably higher than that of RegexT (72.6\%, 95\% CI 63.92-79.94\%).

Limitations\textbf{:} \textbf{\space }Subtle prompt engineering is needed to improve ChatGPT output.\textbf{\space }

Conclusions: Large language models create diverse opportunities and, whilst complex, present an intuitive interface to information but require careful implementation to avoid unpredictable errors.\def\keywordstitle{Keywords}

\medskip\noindent\textbf{\sffamily Keywords: }{obesity, veterinary health informatics, text-mining, Artificial Intelligence, ChatGPT} \\

\noindent\hrulefill
    
\section*{Introduction}
Obesity is a common and significant medical condition in companion animals (1,2). The Small Animal Veterinary Surveillance Network (SAVSNET) collects anonymised electronic health records (EHRs) from veterinary practices in real-time (3). These remain an unexploited resource for investigating canine health, with relevant clinical information often submerged in unstructured free text. Automated systems to surface this information are therefore essential where reading and manual annotation is infeasible. Regular expressions, tools designed to detect fixed word-patterns, have often be used in this setting. Such methods try to try to identify implicit negation (``not vomiting'') and contextual information that indicates a feature is not present (``come back if there are any signs of vomiting'') and require complex rules to accommodate varied/unpredictable language (4). 

Recently, large language models (LLM) including generative pre-trained transformers (such as GPT3.5 which underpins ChatGPT) have become available. These complex neural networks, with hundreds of billions of parameters (5), trained using vast datasets to generate responses to preceding text (4-6) can generate human-like responses to complex prompts. These provide an exciting opportunity for automated data extraction (7-9) and studies assessing their veterinary application are urgently required.

Here we compare the performance of a validated rule-base system using regular expressions (RegexT, 10) to that of a prompt-based approach using ChatGPT to identify body condition score (BCS) of the patient if recorded at the time of the consultation in a sample of publicly available clinical narratives.

\section*{Materials and Methods}
The Small Animal Veterinary Surveillance Network (SAVSNET) collects EHR from a sentinel network of UK veterinary practices. Each EHR contains a clinical narrative written by the attending practitioner. This study used an anonymised  random sample of 4,415 EHRs available in the public domain (11). These narratives were read by domain experts to identify overweight BCSs recorded either on a five-point scale (\ensuremath{\geq }3.5 out of 5 are considered overweight) or nine-point scale (\ensuremath{\geq }6 out of 9 are considered overweight). The collection and use of EHRs by SAVSNET is approved by the Research Ethics Committee at University of Liverpool (RETH001081).

A regular expression (RegexT) was designed to detect overweight BCSs considering the variety of notation used by practitioners to record both the denominator and numerator in free text (Figure 1).

In parallel, we refined a ChatGPT prompt describing four basic rules to coerce output of any BCS along with a prediction regarding overweight status. Data extraction was performed using Python version 3.7.10 (12). GPT3.5 Turbo (13) was accessed through a Python API processing multiple EHRs, each a separate row, appended to the prompt in Figure 2.

The accuracy of these RegexT and ChatGPTs systems were assessed based on the returned BCS using precision (equivalent to positive predictive value), and recall (equivalent to sensitivity). The 95\% confidence intervals (95\% CI) were calculated by modified Wald method using GraphPad (QuickCalcs){\textregistered} (14). 
    
\section*{Results}
\textbf{Extracting overweight BCS measurements with ChatGPT}

Reading 4415 records, identified 117 (2.65\%) reporting an overweight BCS.

ChatGPT identified all 117 overweight BCSs (recall 100\%; 95\%CI 96.18-100\%). The output variably fitted the structure requested by the prompt requiring some manual interpretation. Only 89 were additionally described as being overweight. Twelve GPT outputs appended text either peripheral to the task or generally not relevant to weight such as ``Last WORM and FLEA treatments?'' and ``Possible dietary indiscretion''; some of these additional texts were however relevant (``Coming down in weight. 600g since last September. O~thinks because he is getting less treats.'').

ChatGPT falsely identified BCSs in 14 records; these reflected extraction of other similarly formatted clinical information such as lameness scoring (e.g., EHR-text ``6/10 lameness''; ChatGPT-output ``BCS 6/9. BCS overweight''.) and body weight information (e.g., EHR-text ``wt 6.65kg''; ChatGPT-output: ``BCS 6.65/9''). The precision of ChatGPT was therefore 117/131 (89.3\%; 95\%CI 82.75-93.64\%). 

ChatGPT classified an additional 61 consultations as being overweight in the absence of a recorded overweight BCS. Forty of these were also described as overweight by the attending veterinarian (eg ``she is overweight'', ``normal on clinical exam apart from overweight''). Of the remaining 21 records ChatGPT recorded a normal or low BCS as overweight, or an erroneous false positive high BCS. For example, ChatGPT coded ``BCS 5.75/9 would benefit from further weight loss'' as overweight despite BCS not crossing the threshold of 6. 

ChatGPT failed to return an appropriate answer for 29 records, instead outputting texts such as ``Hello! How can I~assist you today?''. This was often associated with short narratives (27 less than 23 characters in length) such as ``pay now'' and ``all ok sign off'', none containing BCS information.

In comparison, from the 4415 records, RegexT successfully identified 85 of the 117 (recall 72.6\%; 95\% CI 63.92-79.94\%) narratives containing an overweight BCS and no false positives (precision 100\%; 95\% CI 94.81{\textendash}100\%). The 32 overweight BCS missed were associated with format variants not captured by the regex syntax (eg ``BCS: 6-7 out of 9'' and ``BCS: 6/9''). Clearly the regex did not identify any of the 40 narratives that lacked a BCS, but that were described as overweight by the vet and identified by ChatGPT.

Full list of narratives, regex and ChatGPT outputs is available in supplementary material.
    
\section*{Discussion}
To leverage the true value of clinical free text in large health datasets to understand complex diseases like obesity will require careful application of increasingly complex text mining solutions. Regular expressions have been used to identify a wide range of disease phenotypes based on coded patterns of text. More recently, LLMs are offering novel solutions in a wide range of situations. Here we assess the strengths and weaknesses of a LLM (GPT 3.5 Turbo) for a named entity recognition task identifying overweight BCSs in clinical veterinary EHRs, comparing the results to a regular expression and to manual reading. 

Using an intuitive prompt, ChatGPT was coerced to identify both all overweight BCSs and overweight animals without a reported BCS. In contrast, the regex method missed BCSs with novel unpredicted formatting. In this setting of obesity, ChatGPT efficiently identified most overweight animals, with or without a BCS and could be further used to aid the reengineering of systems based on regular expressions. However, occasional false positives were identified by ChatGPT, often associated with other scores like lameness; this behaviour might be avoidable through more subtle prompt engineering (e.g., by prompting: “exclude lameness scores normally recorded out of ten and heart murmur scores, normally recorded out of six”). When high accuracy is the goal, many case studies may still require a final manual classification. Any follow-on manual reading task is made simpler again by prompt engineering, attempting to constrain ChatGPT to tabular output. However, in our study, this frequently failed, making the manual classifying step somewhat more complex than envisaged. Future studies may focus on engineering these prompts to reduce false positives, and to tighten the output structure (15). 

ChatGPT could create overtly false assertions (sometimes described as hallucinations), sometimes comprising understandable unhelpful output. In other settings these can be far more fanciful (16, 17).

\section*{Conclusions}
Despite the complexity of both clinical narratives and the underlying technology of ChatGPT, we were able to efficiently extract overweight BCSs with higher recall than a rule-based system. Challenges remain around the ethics of submitting health texts to an online server; here we used a public domain, anonymised dataset. Issues of cost prohibiting screening bigger datasets may be overcome by installing in-house increasingly available free models. The two systems (regex and language model) offer some complementarity. Future improved prompt engineering will enhance precision and outputted text format. 

\section*{Data Availability Statement}
The data that supports the findings of this study are available in the supplementary material of this article.

\section*{Author contributions}
Hypothesis generation and experimental design: Ivo S. Fins, Sean Farrell, Peter-John Noble. Data analysis and validation: Ivo S. Fins, Heather Davies, Sean Farrell, Gina Pinchbeck, Jose R. Torres, Alan Radford, Peter-John Noble. Writing and revising the manuscript: Ivo S. Fins, Alan Radford, Heather Davies, Sean Farrell, Peter-John Noble.

\section*{Acknowledgements}
Ivo Fins is currently funded by Dogs Trust. We are also grateful for prior support and major funding from BSAVA and BBSRC. Heather Davies is funded by the University of Liverpool and the Veterinary Medicines Directorate (VMD). Sean Farrell is funded as part of a BBSRC-DTP. J Torres is funded by PetPlan Charitable Trust. We wish to thank data providers both in veterinary practice (VetSolutions, Teleos, CVS, and other practitioners) and in veterinary diagnostic laboratories, without whose support and participation this research would not be possible.

\section*{Funding and competing interests statement}
This project was part funded by Dogs Trust (SAVSNET-Agile grant). All authors declare that they have no conflicts of interest.

\section*{References}
1. Chapman M, Woods GRT, Ladha C, Westgarth C, German AJ. An open-label randomised clinical trial to compare the efficacy of dietary caloric restriction and physical activity for weight loss in overweight pet dogs. The Veterinary Journal. 2019 Jan;243:65–73.

2. Chandler M, Cunningham S, Lund EM, Khanna C, Naramore R, Patel A, et al. Obesity and Associated Comorbidities in People and Companion Animals: A One Health Perspective. Journal of Comparative Pathology. 2017 May;156(4):296–309.

3. Sánchez-Vizcaíno F, Jones PH, Menacere T, Heayns B, Wardeh M, Newman J, et al. Small animal disease surveillance. Veterinary Record. 2015 Dec;177(23):591–4.

4. Brown TB, Mann B, Ryder N, Subbiah M, Kaplan J, Dhariwal P, et al. Language Models are Few-Shot Learners. arxivorg [Internet]. 2020 May 28; Available from: https://arxiv.org/abs/2005.14165

5. Vaswani A, Shazeer N, Parmar N, Uszkoreit J, Jones L, Gomez AN, et al. Attention Is All You Need [Internet]. arXiv.org. 2017. Available from: https://arxiv.org/abs/1706.03762 

6. Radford A, Wu J, Child R, Luan D, Amodei D, Sutskever I. Language models are unsupervised multitask learners. OpenAI blog. 2019 Feb 24;1(8):9.

7.  Wang S, Sun X, Li X, Ouyang R, Wu F, Zhang T, et al. GPT-NER: Named Entity Recognition via Large Language Models [Internet]. arXiv.org. 2023. Available from: https://arxiv.org/abs/2304.10428

8. Yan H, Deng B, Li X, Qiu X. TENER: Adapting Transformer Encoder for Named Entity Recognition. arXiv:191104474 [cs] [Internet]. 2019 Dec 10; Available from: https://arxiv.org/abs/1911.04474

9. Hu Y, Ameer I, Zuo X, Peng X, Zhou Y, Li Z, et al. Zero-shot Clinical Entity Recognition using ChatGPT [Internet]. arXiv.org. 2023 [cited 2023 Jul 6]. Available from: https://arxiv.org/abs/2303.16416

10.  Fins I, Bunker L, Radford A, German A, Noble P-J. Tackling canine obesity: development of a regular expression-based tool for uncovering overweight and obese canine patients from veterinary clinical narratives – a pilot study. In Conference Proceedings: Healthcare Text Analytics Conference (HealTAC). UK Healthcare Text Analytics Network. Manchester, 2021.

11. Using SAVSNET data for research - Small Animal Veterinary Surveillance Network (SAVSNET) - University of Liverpool [Internet]. www.liverpool.ac.uk. [accessed 2023 July]. Available from: https://www.liverpool.ac.uk/savsnet/using-savsnet-data-for-research/

12. Python. Python [Internet]. Python.org. Python.org; 2019. Available from: https://www.python.org/

13. ChatGPT [Internet]. openai.com. Available from: https://openai.com/product/chatgpt

14. GraphPad Software [Internet]. www.graphpad.com. [accessed 2023 July]. Available from: http://www.graphpad.com/quickcalcs/

15. Reynolds L, McDonell K. Prompt programming for large language models: Beyond the few-shot paradigm. In Extended Abstracts of the 2021 CHI Conference on Human Factors in Computing Systems 2021 May 8 (pp. 1-7).

16. Roller S, Dinan E, Goyal N, Ju D, Williamson M, Liu Y, et al. Recipes for building an open-domain chatbot. arXiv:200413637 [Internet]. 2020 Apr 30; Available from: https://arxiv.org/abs/2004.13637

17. Maynez J, Narayan S, Bohnet B, McDonald R. On Faithfulness and Factuality in Abstractive Summarization [Internet]. arXiv.org. 2020. Available from: https://arxiv.org/abs/2005.00661

\bibliographystyle{unsrt}

\relax\moveleft-\paperwidth\vbox{\fontsize{.0001pt}{0pt}\selectfont\bibliography{\jobname}}
    
\end{document}